\documentclass[twocolumn,pra,showpacs,final,superscriptaddress,10pt,aps,floatfix]{revtex4-1}
\usepackage[usenames]{color}
\usepackage{amsmath}
\usepackage{amssymb}
\usepackage{mathrsfs}
\usepackage{rotating}
\usepackage{bbold}
\usepackage{euscript}
\usepackage{graphicx}
\usepackage{graphics}
\usepackage{bm}
\begin{document}



\title{Resonant and nonresonant for population transfer by impulsive X-ray stimulated Raman scattering}

\title{Optimizing impulsive X-ray Raman scattering for population transfer in atomic systems}

\author{James P. Cryan}
\affiliation{Stanford PULSE Institute for Ultrafast Energy Science, SLAC National Accelerator Laboratory, Menlo Park, CA 94025}
\author{Matthew R. Ware}
\affiliation{Stanford PULSE Institute for Ultrafast Energy Science, SLAC National Accelerator Laboratory, Menlo Park, CA 94025}
\author{Daniel  J. Haxton}
\affiliation{Department of Physics, University of California, Berkeley CA 94720}


\begin{abstract}

Impulsive X-ray Raman excitations of Lithium, Neon, and Sodium are calculated using the Multiconfiguration 
Time-Dependent Hartree-Fock method.  Using linearly polarized laser pulses without chirp,
we determine the optimum central frequency, intensity, and duration for maximum population transfer to valence excited 
states.  
%
%
We demonstrate the existence of
two local optima or ``sweet spots'' for population transfer, either of which, depending on the system, may be superior.  For some
systems we find that population transfer can be maximized by nonresonant Raman transitions, 
red-detuned below K-edge, because such detuning minimizes core-excited populations and ionization loss. For instance,
in Neon near the K-edge the global optimum for population transfer occurs at 
high intensity (8 $\times$ 10$^{19}$ W cm$^{-2}$), short duration (82as full-width-at-half-maximum), 
and 24eV red-detuned from the K-edge.  

\end{abstract}

\maketitle

\section{Introduction}

The advent of high intensity broad bandwidth x-ray sources has driven the extension of non-linear 
spectroscopic techniques, developed for infrared and visible wavelengths, into the x-ray regime. 
For example, stimulated x-ray Raman scattering (SXRS) has been suggested as a technique to 
prepare and probe coherent superpositions of valence excited electronic 
states~\cite{TanakaPRA2003,biggs_two-dimensional_2012,biggs_watching_2013,mukamel2013}.
Despite these suggestions,
there have been few calculations of population transfer with broad bandwidth x-ray pulses
beyond perturbative models.
To this end, we have applied our implementation of Multiconfiguration Time-dependent Hartree-Fock 
(MCTDHF)~\cite{prolate, restricted, sincdvr} to predict valence excited state population transfer resulting from
impulsive SXRS in atomic Lithium, Neon, and Sodium.

We have tried to maximize the population transfer to particular valence excited states in these
atoms.  The maximum population transfer is shown to be obtained at intensities much higher
than that at which 2nd order perturbation theory is valid.  Higher-order effects, in particular AC
stark shifts, substantially modify the result that would be predicted from 2nd order perturbation theory.

In general, we find two ``sweet spots'' for population transfer to valence excited states.  These are different local
maxima for population transfer as a function of central frequency, intensity, and duration (bandwidth)
of the pulse.  We consider linearly polarized fields without chirp only for this study.

The first sweet spot, or local maximum for population transfer, is that which has been studied by previous authors~\cite{tiger}.
For a single discrete transition driving the impulsive Raman process near this local optimum, 
the central frequency is such that the
pump and Stokes transitions are red- and blue-detuned, respectively.  
We find well-defined local maxima for population transfer in these three systems near the K-edge, at this conventional-wisdom
sweet spot.  However, in contrast to prior studies we show that
continuum oscillator strength is generally important for driving the transition near this optimum, and in none of the
cases studied here does a simple 3-state model seem to suffice even at 2nd order.

The second sweet spot often appears and is found here to be dominant in transferring population
in the Neon atom.  This sweet spot occurs at shorter durations, higher intensities
(greater than 10$^{17}$ W cm$^{-2}$) and well red-detuned from the K-edge and any of the near-edge fine structure.

The mechanism behind this red-detuned, higher-intensity, shorter-duration sweet spot would appear to be nonresonant 
Raman via the coherent combination of K-edge continuum and near-edge fine structure.  As the duration of the pulse
decreases the contribution of the near-edge fine structure is lessened and the Raman transition becomes entirely
driven by the continuum oscillator strength.  The red detuning makes the transition nonresonant, and the
magnitude of the Raman wave function (the first-order wave function, corresponding to the core excited intermediate
state) is minimized.  Transitions down from the core-excited intermediates 
to the desired valence states are preferred, whereas the absorption of an additional photon 
is suppressed by the red detuning.


\begin{figure}
\begin{tabular}{c}
\resizebox{0.6\columnwidth}{!}{\includegraphics*[0.6in,0.6in][5.9in,4.2in]{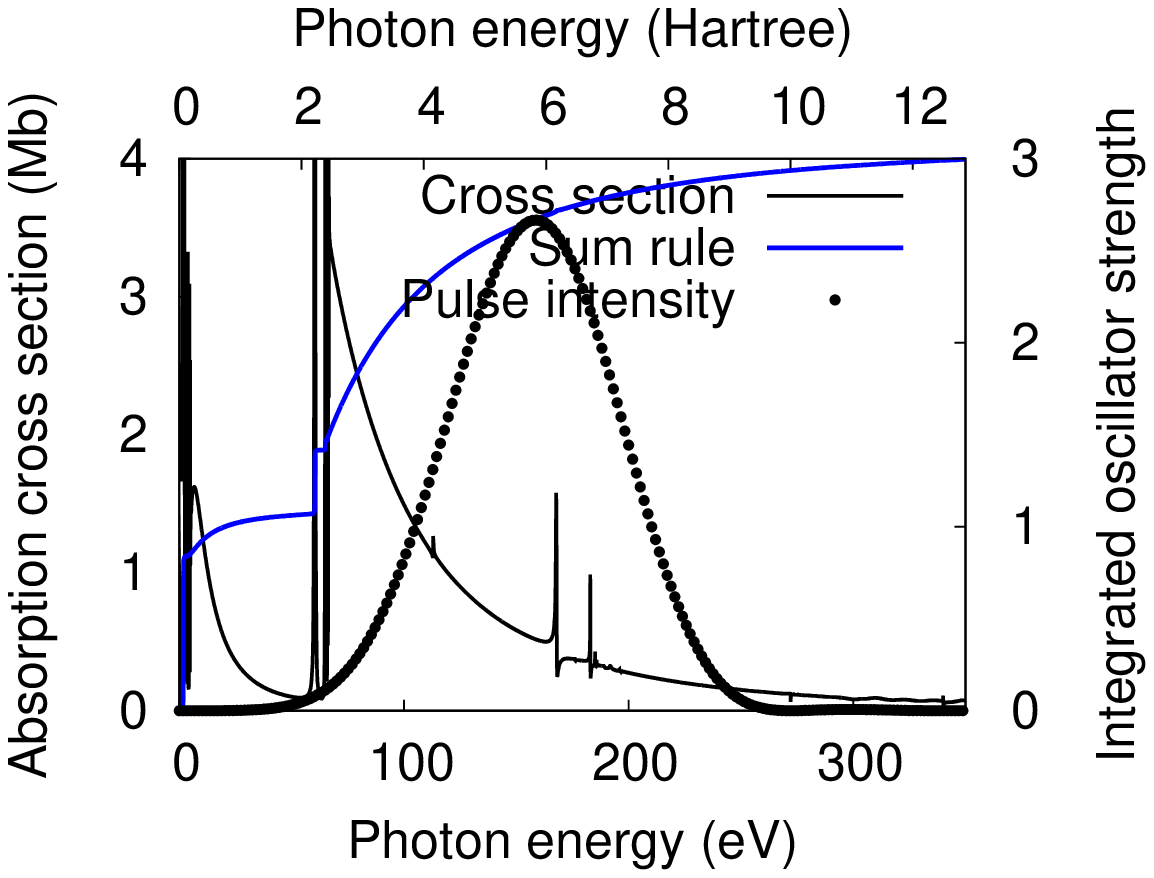}} \\
\resizebox{0.6\columnwidth}{!}{\includegraphics*[0.6in,0.6in][5.9in,4.2in]{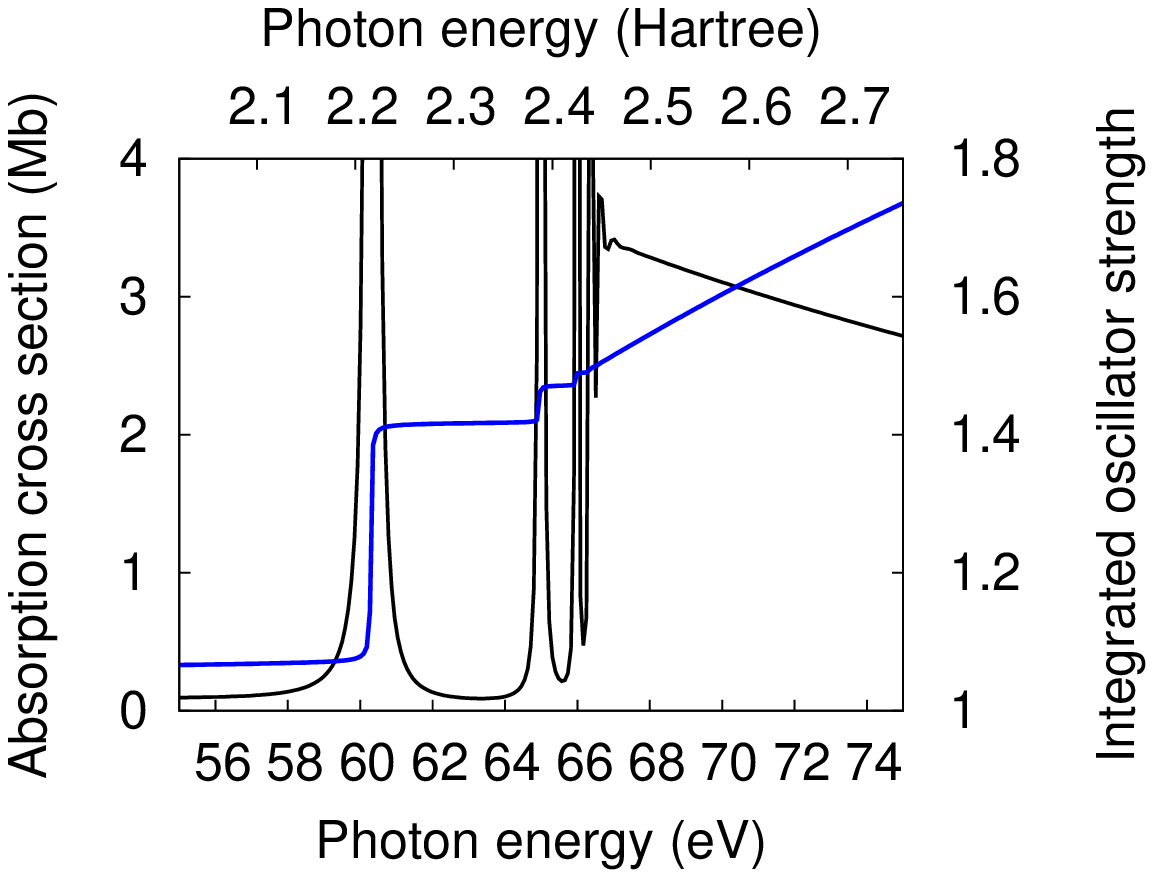}}
\end{tabular}
\caption{(Color online) Photoabsorption cross section and the Thomas-Kuhn sum rule for the lithium atom.
\label{lithium_xfig}}
\end{figure}

\begin{figure*}
\begin{tabular}{ccc}
\resizebox{0.6\columnwidth}{!}{\includegraphics*[0.6in,1.1in][5.9in,4.2in]{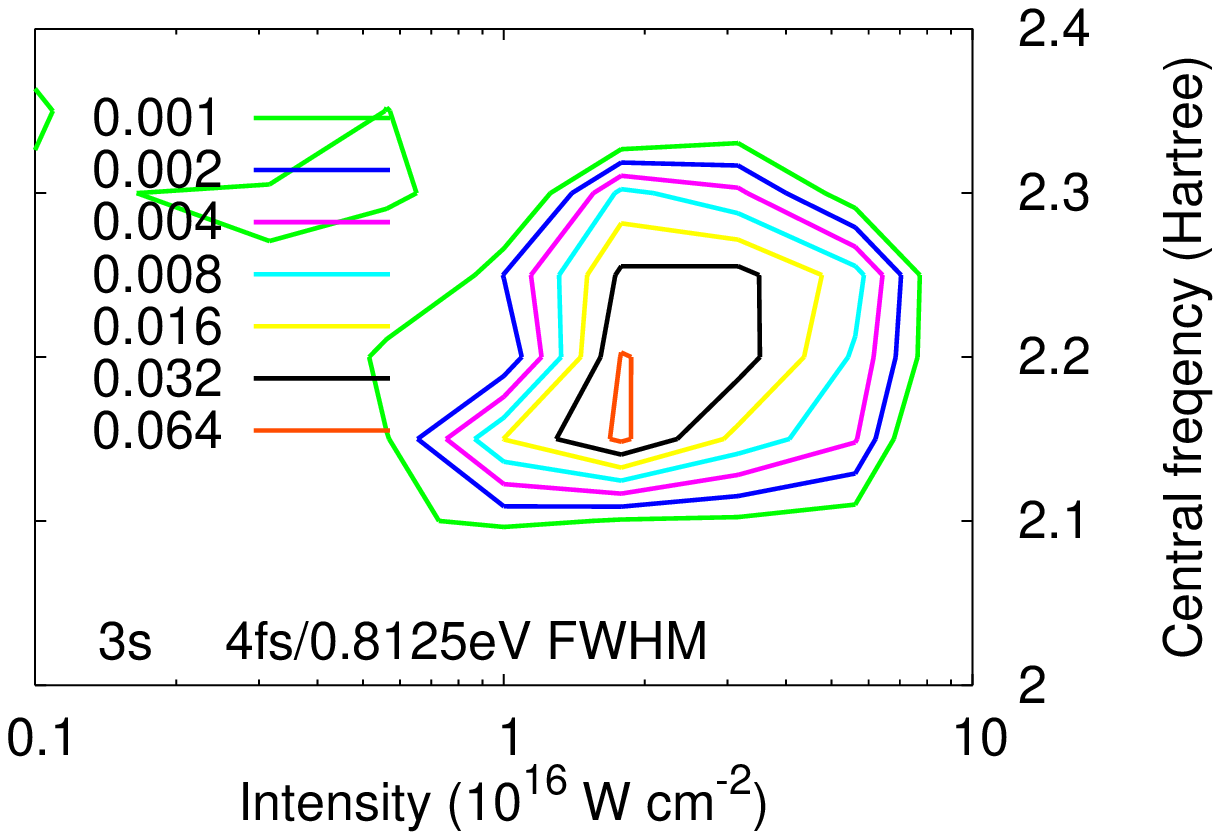}} &
\resizebox{0.6\columnwidth}{!}{\includegraphics*[0.6in,1.1in][5.9in,4.2in]{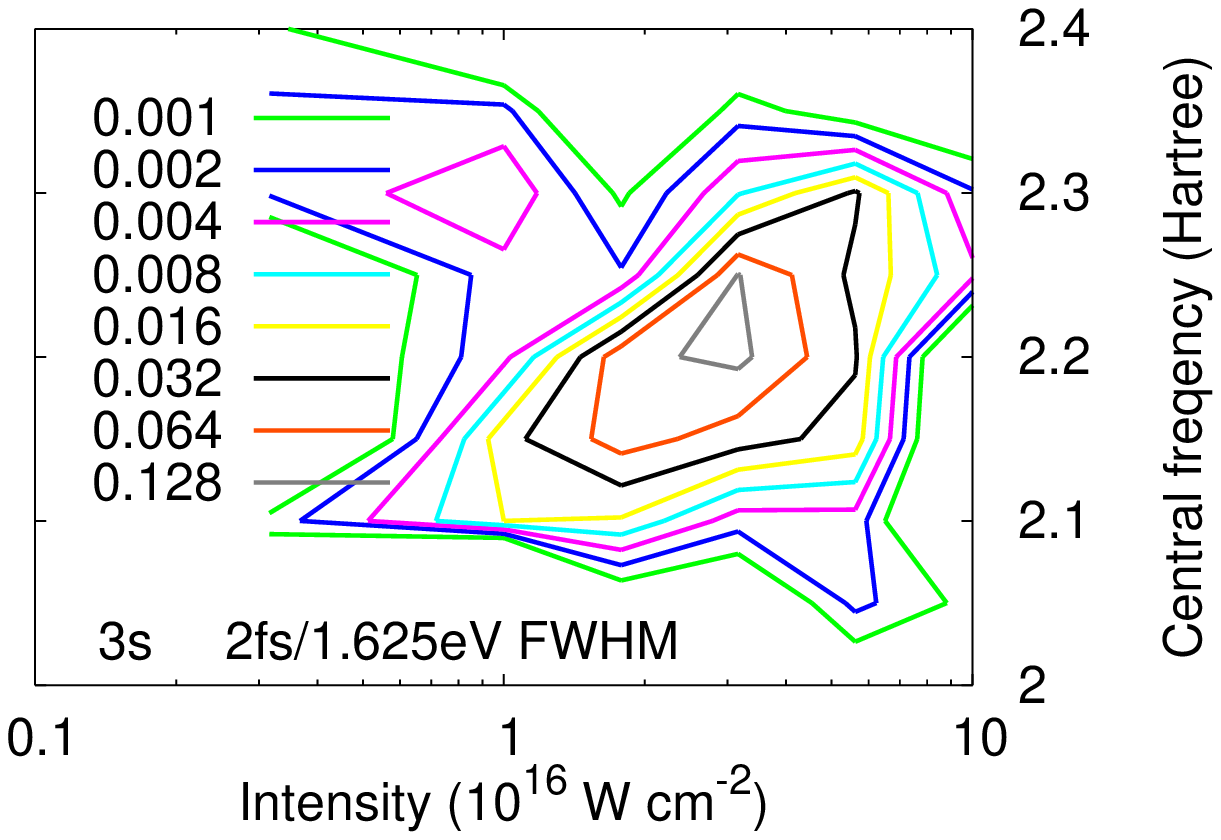}} &
\resizebox{0.6\columnwidth}{!}{\includegraphics*[0.6in,1.1in][5.9in,4.2in]{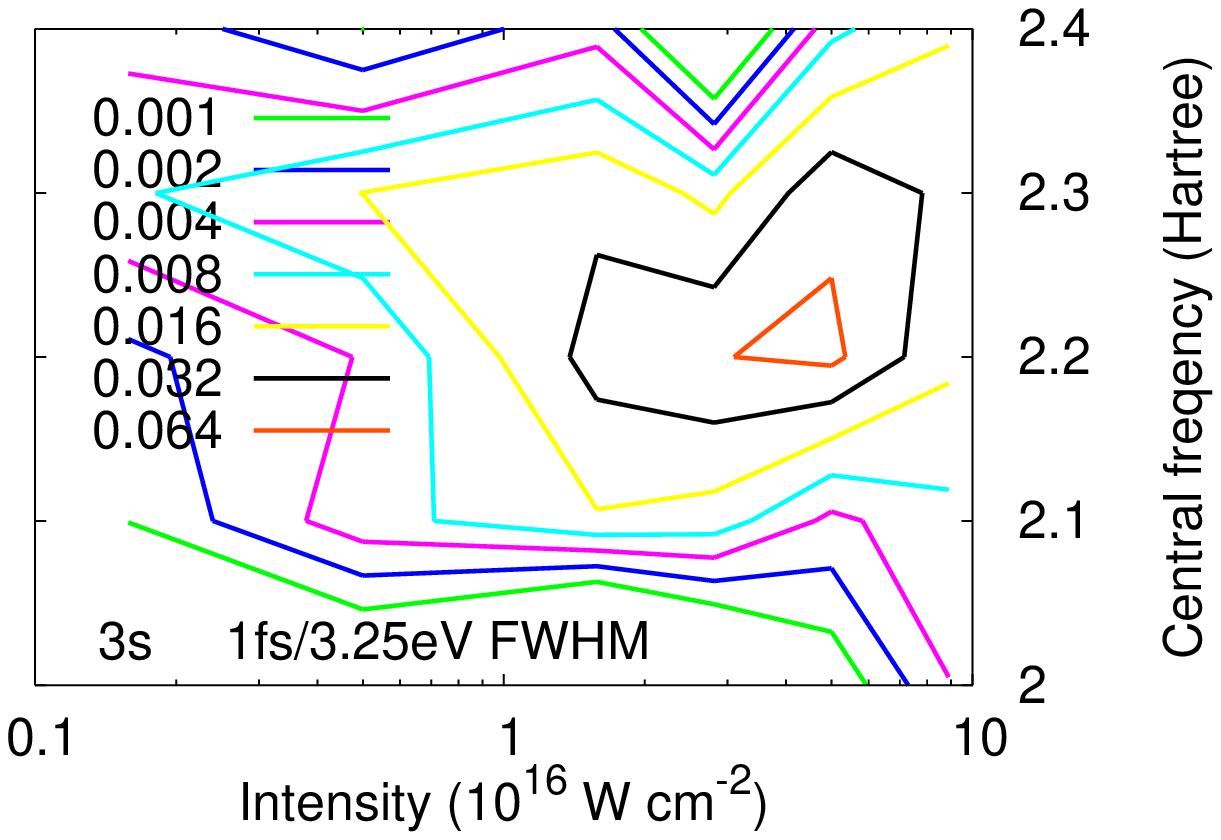}} \\
\resizebox{0.6\columnwidth}{!}{\includegraphics*[0.6in,0.6in][5.9in,4.2in]{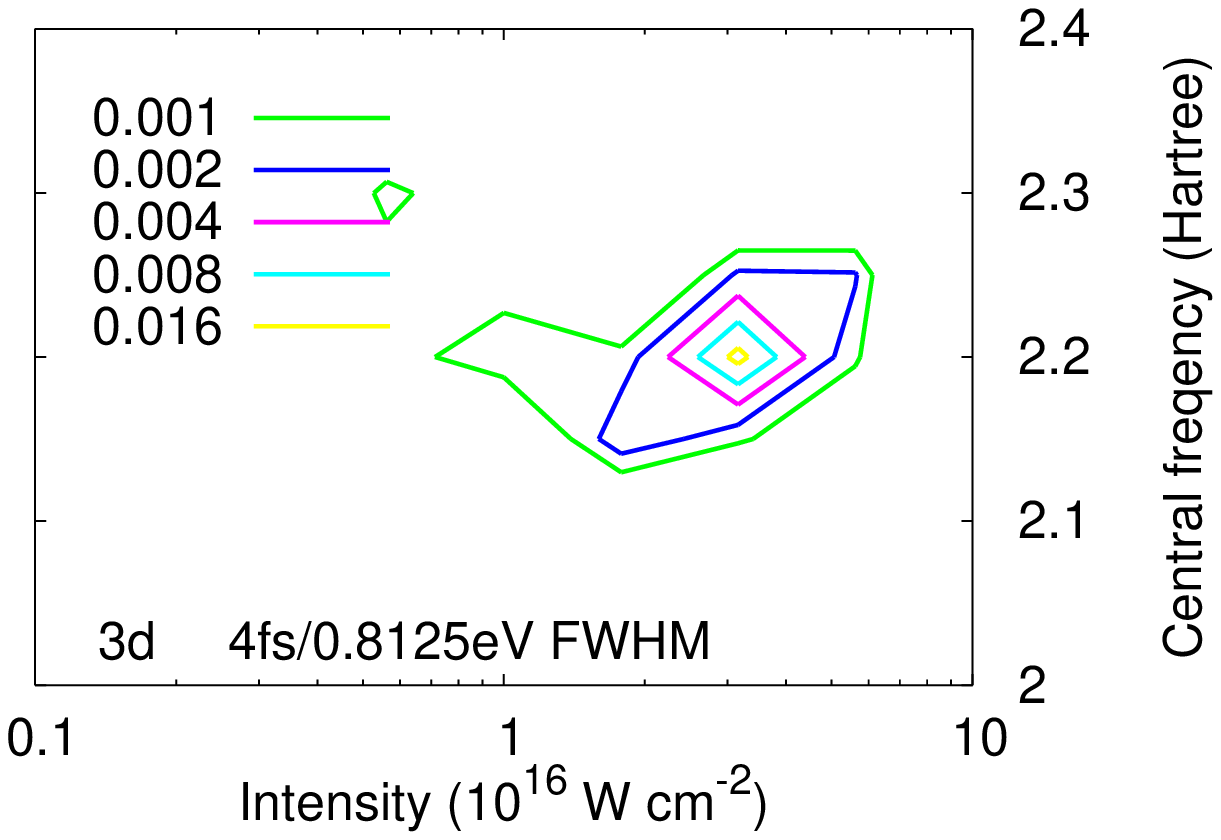}} &
\resizebox{0.6\columnwidth}{!}{\includegraphics*[0.6in,0.6in][5.9in,4.2in]{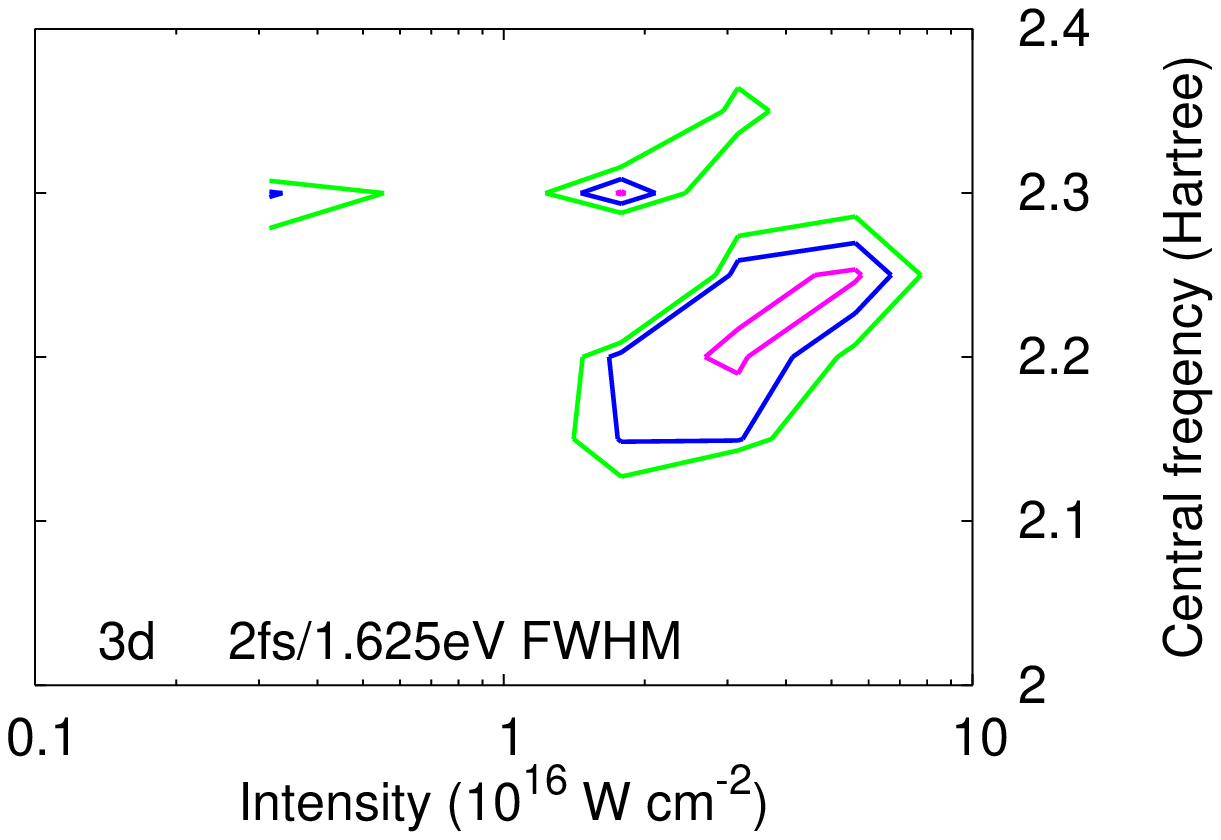}} &
\resizebox{0.6\columnwidth}{!}{\includegraphics*[0.6in,0.6in][5.9in,4.2in]{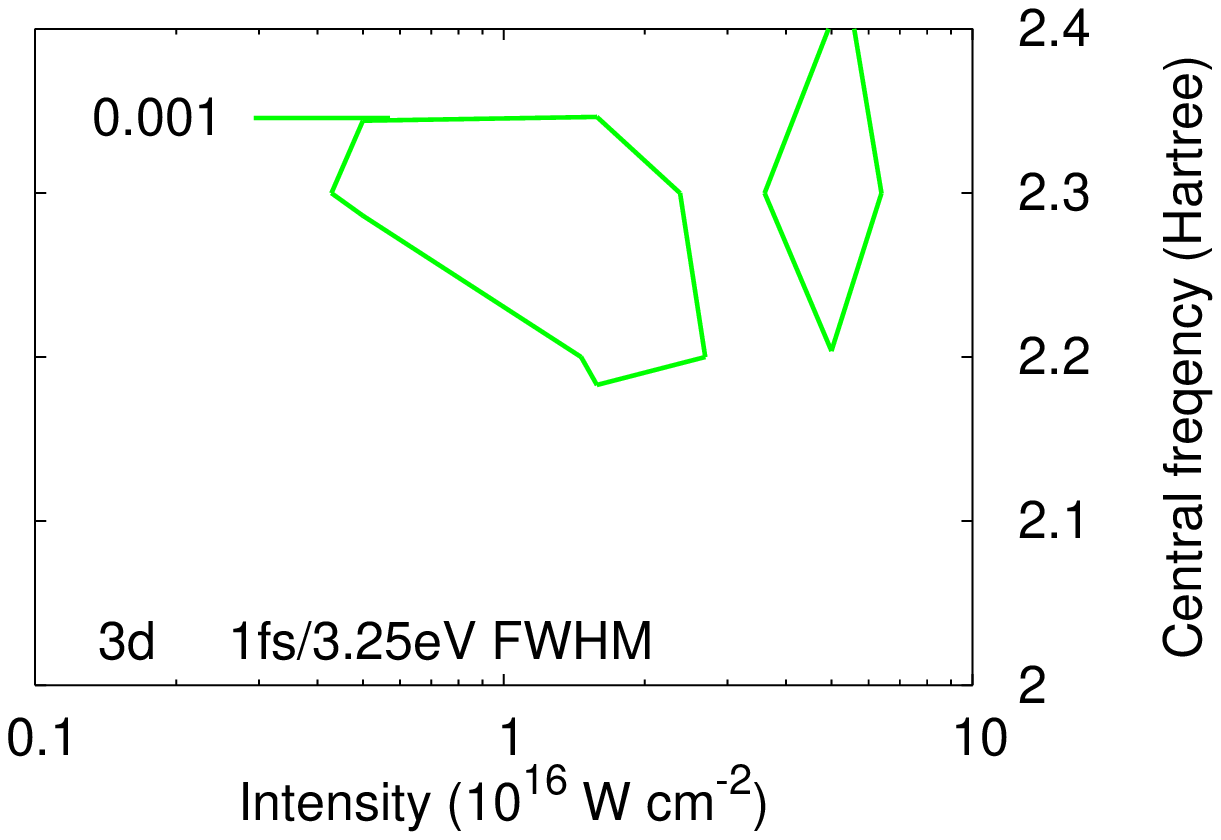}} \\
\end{tabular}
\caption{(Color online) Populations of Lithium 3s and 3d final states.
\label{lithium_popfig}}
\end{figure*}

\section{MCTDHF calculation of impulsive Raman in atoms}

Our implementation~\cite{lbnl_amo_mctdhf} of MCTDHF for electrons in molecules has already been 
described~\cite{prolate, restricted, sincdvr}
Briefly, the MCTDHF method solves the time-dependent Schrodinger
equation using a time-dependent linear combination of Slater determinants, in which the Slater determinants are comprised
of time-dependent orbitals.  The working equations are obtained through
application of the Lagrangian variational principle~\cite{broeck, ohta} to this wave function ansatz.

We employ a pulse in the dipole approximation with central frequency $\omega$ and duration $\pi / \Omega$ as follows.
In the velocity gauge we define the vector potential
\begin{equation}
A_z(t) = \sin^2(\Omega t) \sin\left(\omega \left[t - \frac{\pi}{2 \Omega}\right] \right) \quad (0 \le t \le \frac{\pi}{\Omega})
\end{equation}
In the length gauge we employ the electric field $E_z(t) = \frac{\partial}{\partial t} A_z(t)$.  The results are gauge-independent
on the scale of the figures presented here.
For such a pulse, 
the full width at half-maximum (FWHM) of $E_z(t)$ multiplied by the FWHM of its squared Fourier transform $\vert E_z(\omega)\vert^2$
is 3.25eV-femtoseconds, compared with 2.67eV-femtoseconds for a transform-limited (Gaussian) pulse.  We quote the FWHM of the
envelope of the laser pulse as a function of time, and the FWHM of its squared Fourier transform (spectral density), in energy or frequency.
Our 1fs FWHM pulse, 
with full duration 2fs, has a FWHM of 3.25eV; our 500as FWHM pulse has FWHM 6.5eV; etc.

\section{Lithium results}

The results we have calculated for the Lithium atom indicate that for this species, population transfer via impulsive Raman
transitions is best driven through discrete transitions, in agreement with the conventional wisdom.

Lithium calculations are performed with 15 orbitals, including the 1$s$ and 2$s$ orbitals that are doubly and singly
occupied in the Hartree-Fock configuration, additionally
with two $s$, two $p$ shells, and one $d$ shell of orbitals that 
mostly describe correlation between the 1$s$ orbitals in the initial state.  Once the pulse is applied the symmetry labels are
no longer valid but the orbitals retain their projection of angular momentum on the $z$ axis.
The photoabsorption cross section calculated for the Lithium atom using a 50 femtosecond wave function propagation
and a very weak (10$^{-11}$ W cm$^{-2}$) broadband pulse
is shown in Fig.~\ref{lithium_xfig}.  The lower panel shows the
region near the 1$s$ edge.  The two states with configuration $1s^12s2p$ are not distinguished in this result, which was obtained
after 50fs propagation.  They appear as a single peak at about 60eV (2.2 Hartree). Their combined oscillator strength can be seen to 
be about 0.34.

However, despite the large oscillator strength carried by these $1s \rightarrow 2p$ transitions, 
it is clear that the population transfer driven by impulsive
X-ray Raman excitation involves more than just these transitions.  Results for population transfer to 3$s$ and 3$d$ are
shown in Fig.~\ref{lithium_popfig}.  On the left-hand side of the figures we can see the second-order behavior at lower intensity.
The maximum with respect to central frequency at low intensity is seen to be about 2.3 Hartree in these figures.  The transition
energy of 2.3 Hartree lies in-between the strong $1s \rightarrow 2p$ transition at 2.2 Hartree, and the remaining oscillator strength
including the $1s \rightarrow 3p$ transition at about 2.4 Hartree and the apparent edge at 2.45 Hartree, as seen in 
Fig.~\ref{lithium_xfig}.  The optimum central frequency of 2.3 Hartree implies that the optimum population transfer at low 
intensity is driven by a linear combination of the Rydberg excitations, and also possibly the 1$s$ continuum.  

Furthermore, it is clear that higher-order effects are important for describing the optimum population transfer in the lithium atom.
In a three-state model, assuming a transform-limited pulse, at constant fluence,
second-order perturbation theory predicts that the optimum population transfer occurs when the full width at half maximum (FWHM)
of the spectral intensity is approximately 118\% ($\sqrt{2 \log 2} = 1.17741$)  of the transition frequency.  The experimental value
for the transition energy $2s \rightarrow 3s$ is 3.37eV~\cite{nist}.  From second-order perturbation theory we expect optimum
population transfer with a pulse having 3.97eV FWHM. 

However, the optimum population transfer occurs for pulses with a significantly longer duration and smaller bandwidth.  The optimum
apparent in the results in Fig.~\ref{lithium_popfig} occurs at central frequency about 2.22 Hartree, intensity about 3$\times$ 10$^{16}$
W cm$^{-2}$, for the pulse with FWHM 2fs or 1.625eV, a much longer duration or smaller bandwidth than would be expected.  The
simplest explanation seems to be an AC stark shift of the transition energy, a second-order effect on its own, suggesting that this
behavior might be reproduced at overall fourth order.
We will pursue further analysis of the impulsive Raman excitation in Lithium in subsequent work.

\section{Sodium K-edge results}

Our results on Sodium indicate that both near-edge optimum for population transfer, and also the red-detuned, high-intensity
local optimum may provide good population transfer in Sodium near the 1$s$ K-edge.  
We find that optimum population transfer is driven by continuum transitions for both local optima.

\begin{figure}
\begin{tabular}{c}
\resizebox{0.8\columnwidth}{!}{\includegraphics*[0.7in,0.6in][5.6in,4.2in]{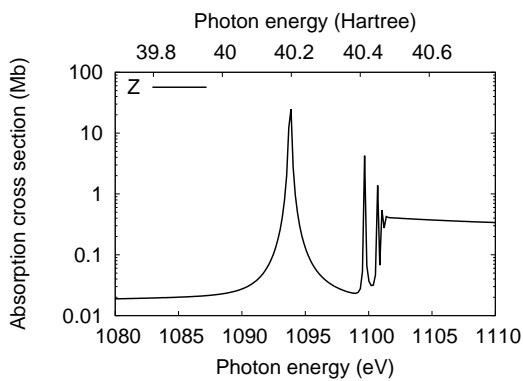}} \\
\end{tabular}
\caption{(Color online) Photoabsorption cross section of Sodium calculated in the vicinity of the 1s edge.
\label{sodium_xfig}}
\end{figure}

We calculate wave functions for Sodium using 10 orbitals.  In the initial state these orbitals are the six Hartree-Fock
orbitals, plus an additional $s$ and $p$ shell which mostly describe correlation within the doubly occupied n=1 and n=2
shells.  The photoabsorption cross section is shown in Fig.~\ref{sodium_xfig}.  The prominent $1s \rightarrow 3p$ excitation
occurs at 40.2 Hartree or 1094eV, and the subsequent members of this Rydberg series converge to the 1s K-edge which is apparent
at about 40.45 Hartree or 1102eV.

\begin{figure}
\begin{tabular}{c}
\resizebox{0.6\columnwidth}{!}{\includegraphics*[0.8in,0.4in][5.7in,4.0in]{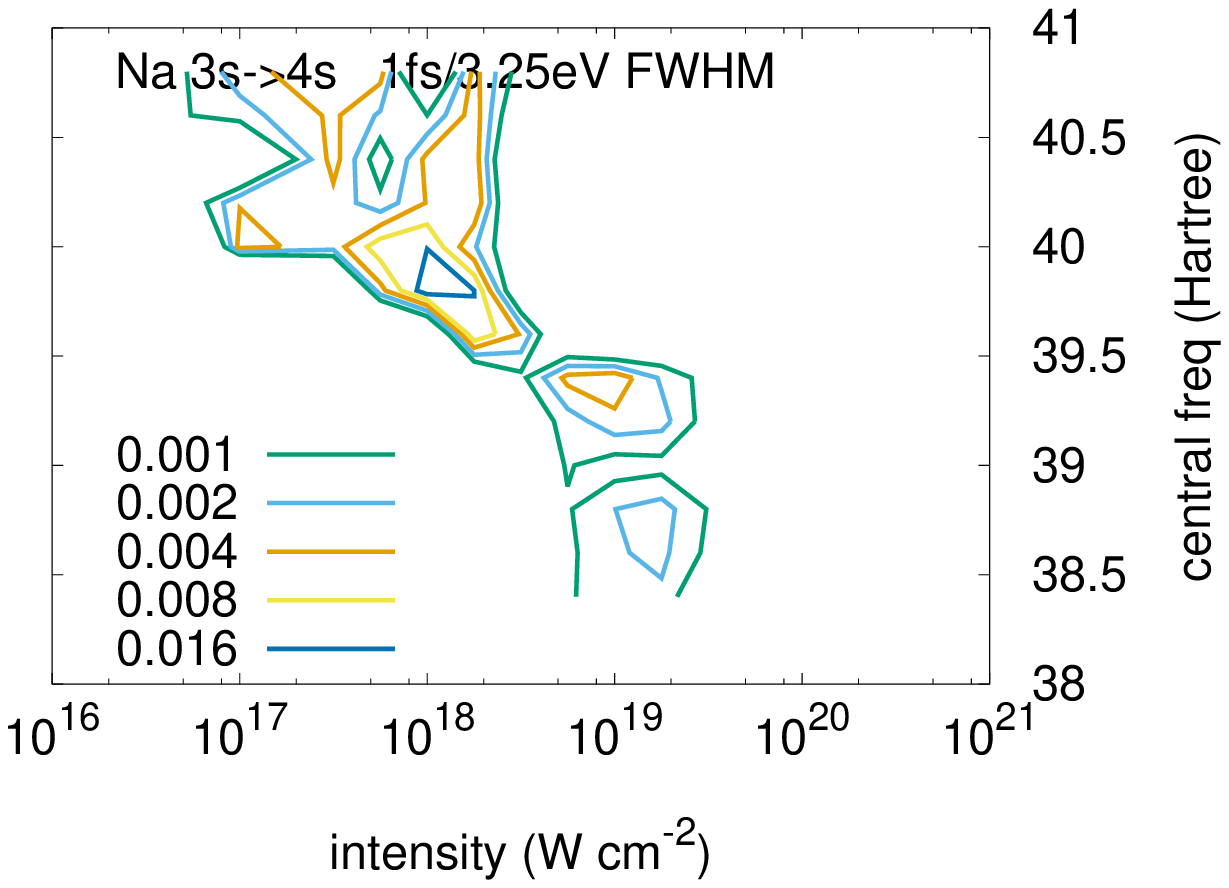}} \\
\resizebox{0.6\columnwidth}{!}{\includegraphics*[0.8in,0.4in][5.7in,4.0in]{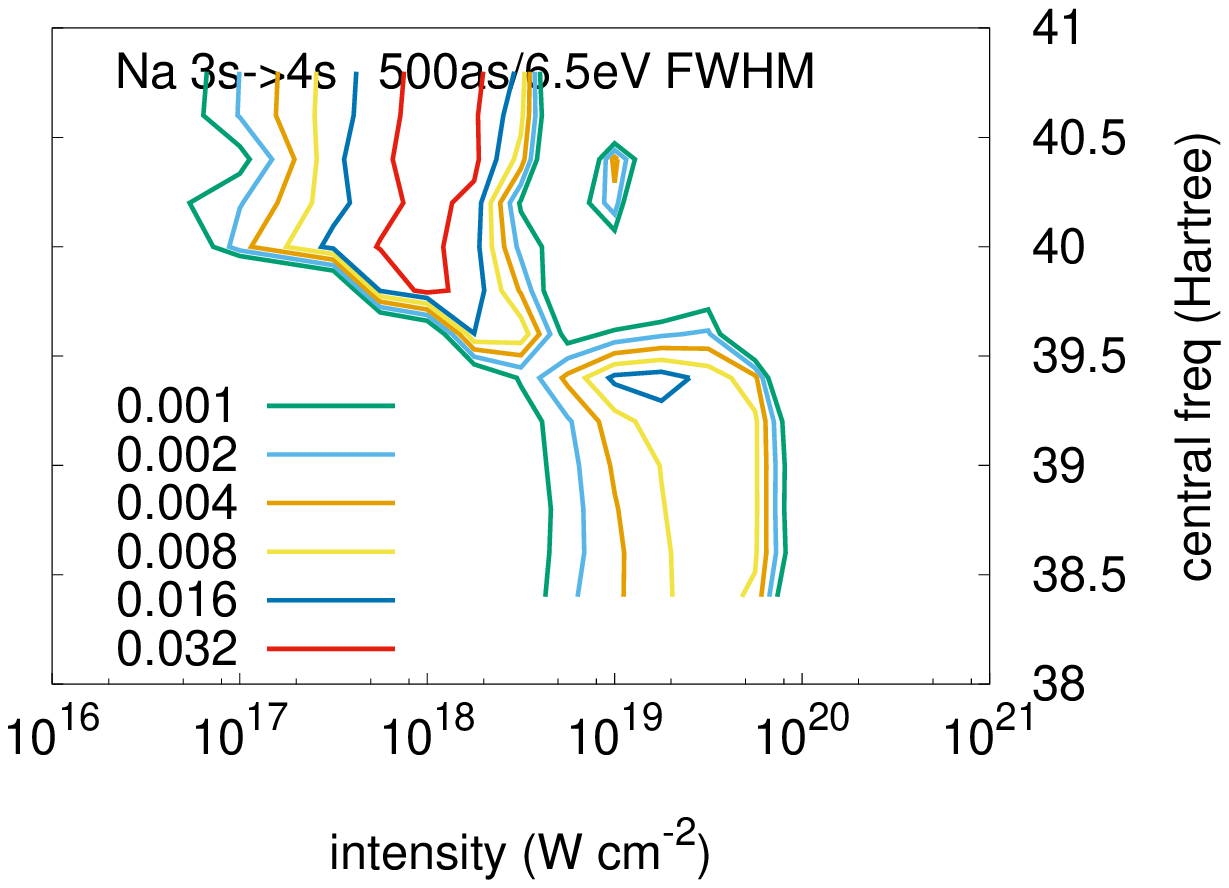}} \\
\resizebox{0.6\columnwidth}{!}{\includegraphics*[0.8in,0.4in][5.7in,4.0in]{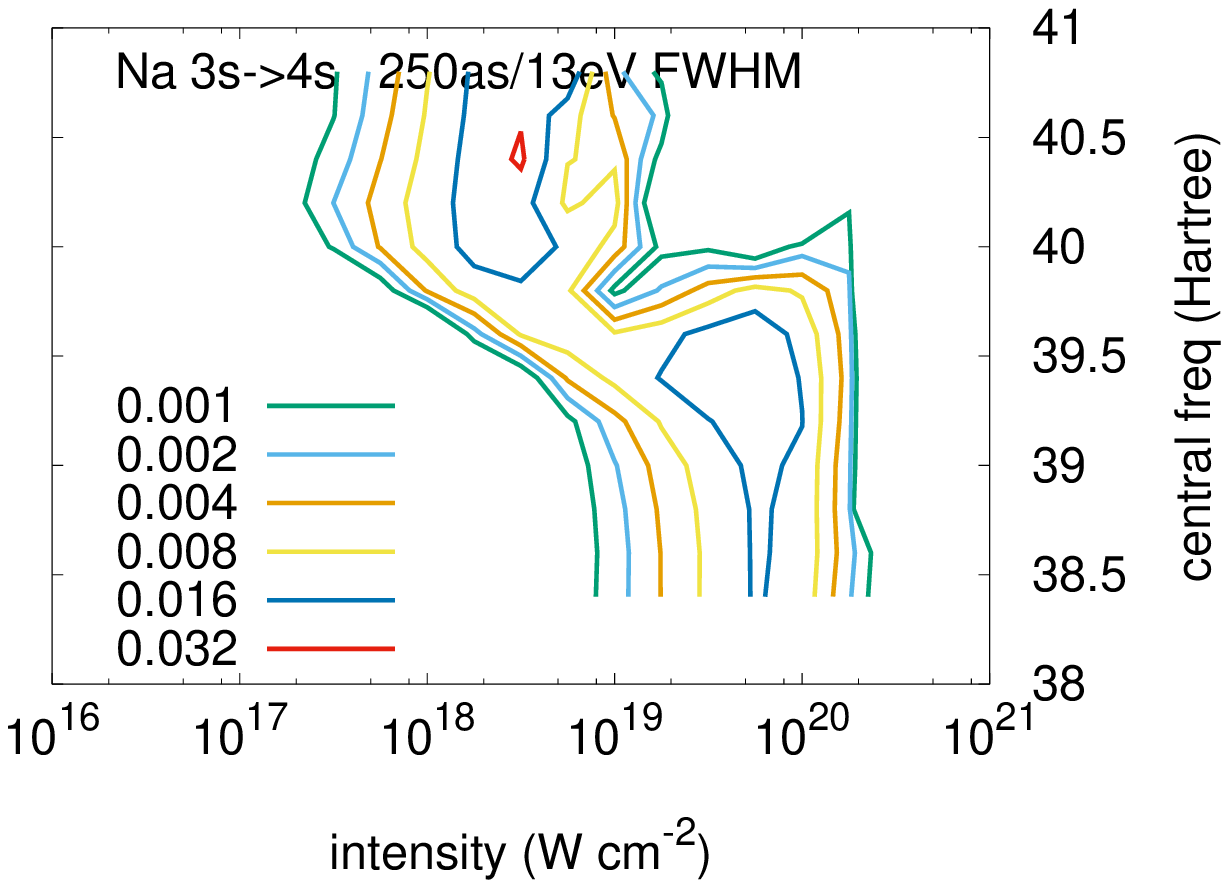}} \\
\resizebox{0.6\columnwidth}{!}{\includegraphics*[0.8in,0.4in][5.7in,4.0in]{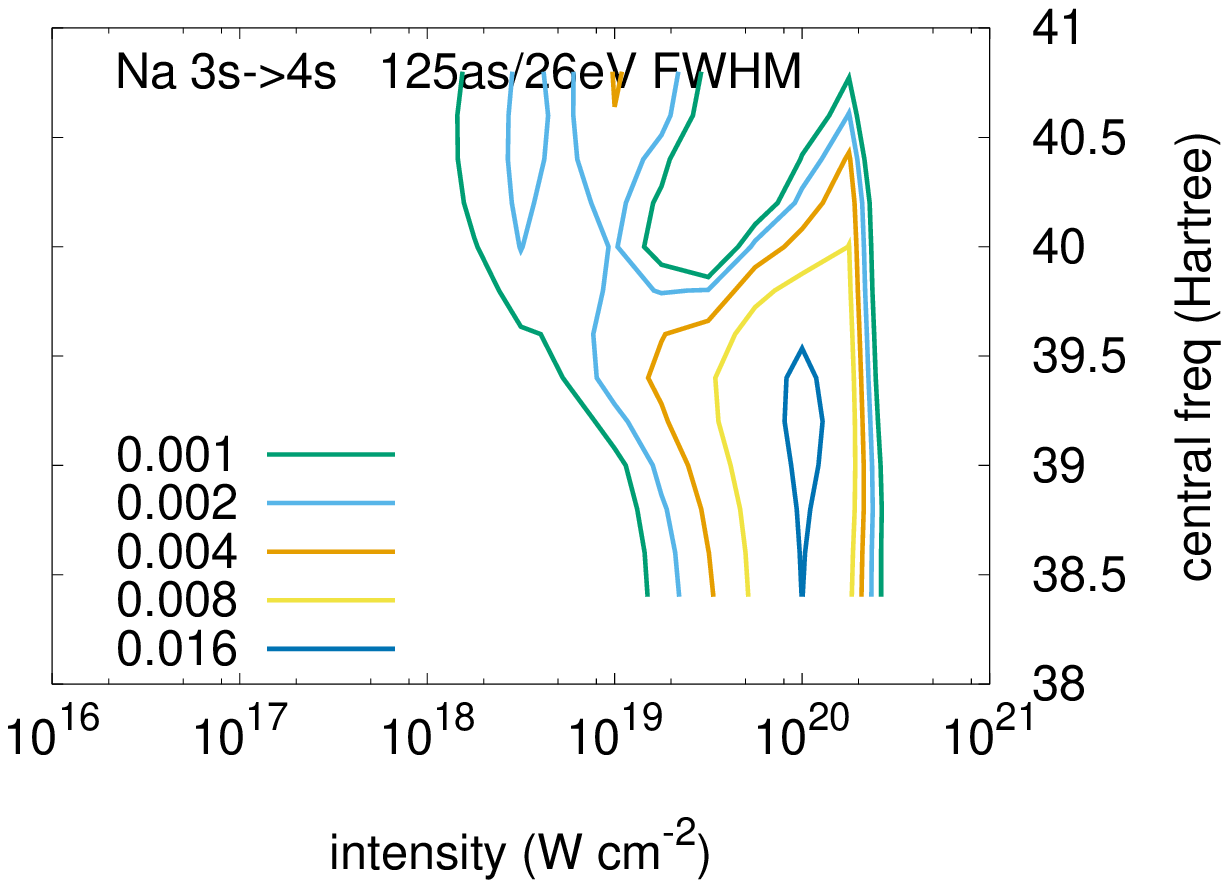}} \\
\end{tabular}
\caption{(Color online) Populations calculated for the 4s state in Sodium.
\label{sodium_popfig}}
\end{figure}

The populations obtained for the 4s state in Sodium are shown in Fig.~\ref{sodium_popfig}.  These results for population transfer
clearly demonstrate the existence of two local maxima, two sweet spots.  The conventional-wisdom sweet-spot occurs around 10$^{18}$
W cm$^{-2}$ at energies near and above the 1$s$ K-edge.  The red-detuned sweet spot occurs around 10$^{19}$ to 10$^{20}$
W cm$^{-2}$, about one hartree (27.2114eV) below the K-edge.  In terms of fluence, the red-detuned sweet spot occurs reliably at about
10,000 J cm$^{-2}$ regardless of duration.
Structure corresponding 
to the discrete transitions is only clearly apparent at the
longest duration, 1fs/3.25eV FWHM.  The 500as/6.5eV FWHM results show little evidence of discrete transitions and the shorter-duration
results show none.  

By fitting the results to a quadratic function of fluence and bandwidth, we obtain the following positions and magnitudes for the two sweet
spots for population transfer via impulsive Raman transitions near the 1s edge in Sodium.  These fits are approximate given the large spacing
of the data points that we have computed, and the somewhat-arbitrary selection of a few points near the apparent maximum.  
The conventional-wisdom sweet spot is found to occur at about 8eV FWHM (406as FWHM) at 40.55 Hartree (1103.4eV), very near the actual
edge, intensity 1.8 $\times$ 10$^{18}$ W cm$^{-2}$, with a maximum population transfer of 4.2\%.  
We note that this conventional-wisdom 
sweet spot is clearly extended and good population transfer may be obtained just below 40 Hartree, extending above 41 Hartree.
The large extent of this sweet spot indicates that it is driven by the K-edge continuum oscillator strength, not discrete transitions.

By fitting the results that we have computed, the red-detuned sweet spot is found to occur at 14.2eV FWHM (228as FWHM), 39.3 Hartree
(1069.4eV), intensity 6.2 $\times$ 10$^{19}$ W cm$^{-2}$, with population transfer of 2.8\%.

In summary, for Sodium K-edge, we find 4.2\% population transfer for the conventional-wisdom sweet spot, and 2.8\% population transfer
via the red-detuned sweet spot.  The conventional-wisdom sweet spot is dominant, but both sweet spots provide good population transfer
of a similar magnitude.

\section{Neon K-edge results}

\begin{figure}
\begin{tabular}{c}
\resizebox{0.8\columnwidth}{!}{\includegraphics*[0.7in,0.6in][5.6in,4.2in]{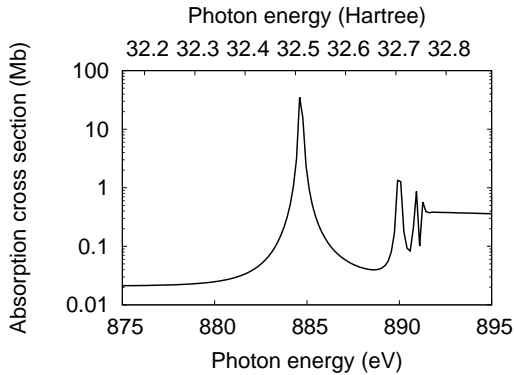}} \\
\end{tabular}
\caption{(Color online) Photoabsorption cross section of Neon calculated in the vicinity of the 1s edge.
\label{neon_xfig}}
\end{figure}

For impulsive Raman transitions in Neon near the 1s K-edge, we find that the red-detuned sweet spot is dominant for population
transfer via impulsive Raman.  The results on Sodium indicated that the red-detuned sweet spot occurs at shorter duration, larger bandwidth.
Since the valence excitation energies in Neon are greater than they are in Sodium, the population transfer occurs with shorter-duration
pulses, and we speculate that short duration pulses and high excitation energies favor the red-detuned sweet spot in general.

We calculate wave functions of Neon using 14 orbitals, the n=1 and 2 shells that are occupied in the Hartree-Fock
configuration plus an additional n=3 shell for correlation.  The photoabsorption cross section is shown in Fig.~\ref{neon_xfig}.
The $1s \rightarrow np$ Rydberg is apparent including the $1s \rightarrow 2p$ excitation at about 885eV or 32.5 Hartree.
It converges to the 1s K-edge at about 892eV or 32.75 Hartree.

\begin{figure}
\begin{tabular}{c}
\resizebox{0.6\columnwidth}{!}{\includegraphics*[0.8in,0.4in][5.7in,4.0in]{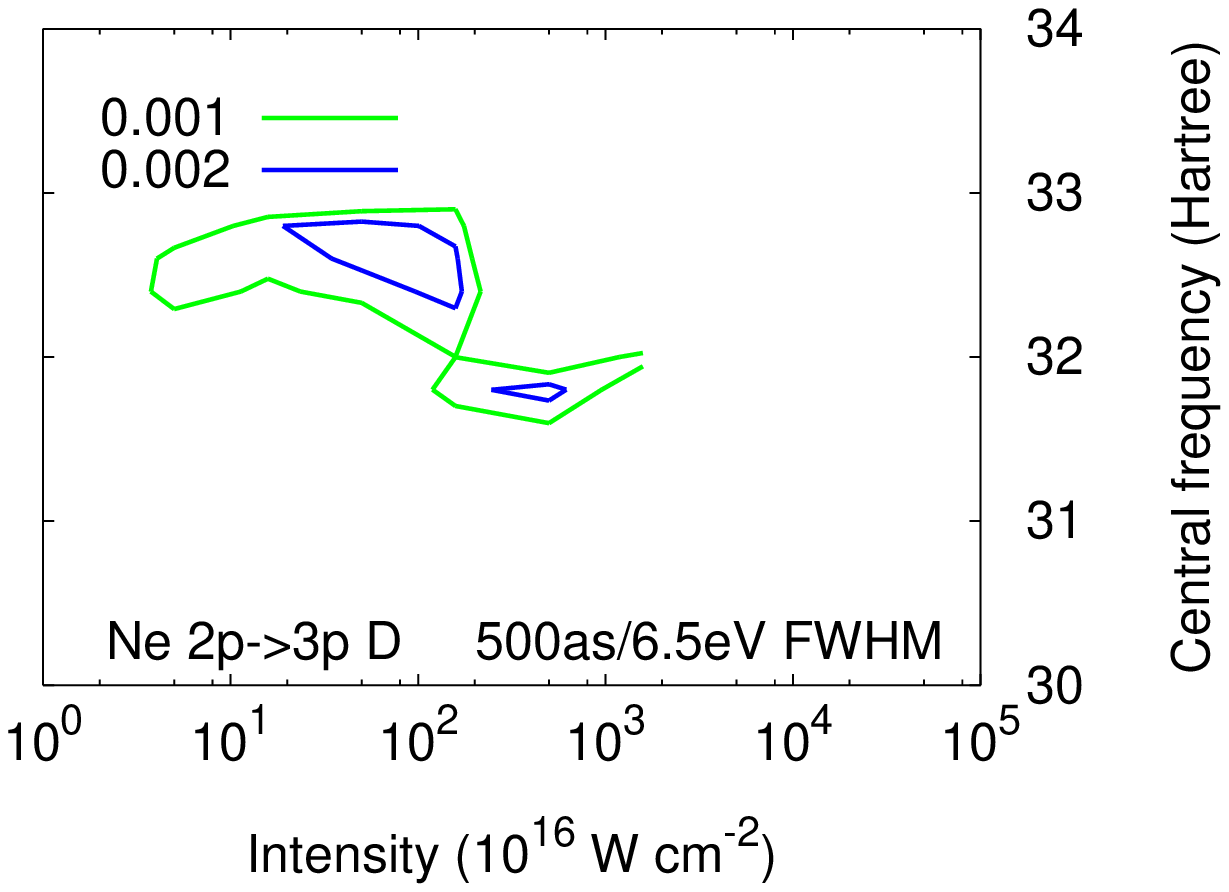}} \\
\resizebox{0.6\columnwidth}{!}{\includegraphics*[0.8in,0.4in][5.7in,4.0in]{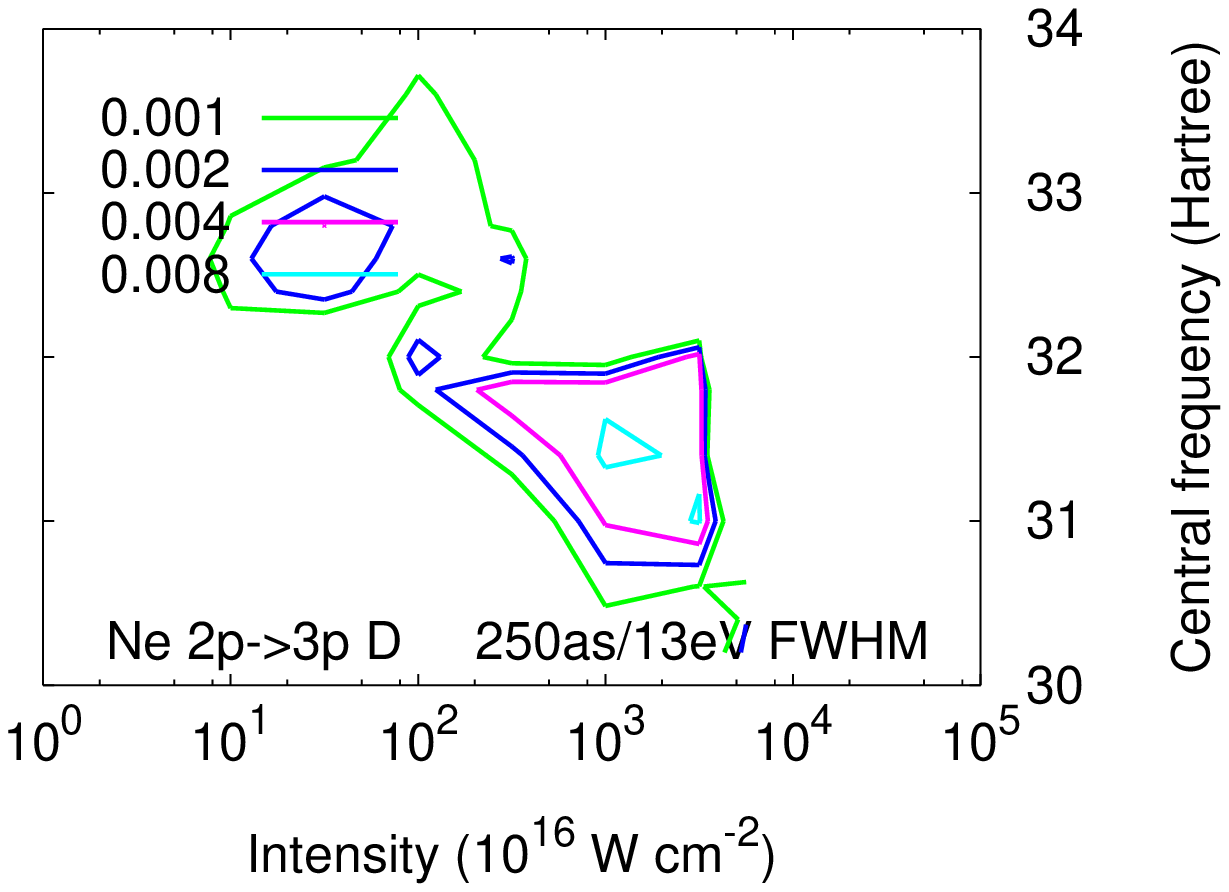}} \\
\resizebox{0.6\columnwidth}{!}{\includegraphics*[0.8in,0.4in][5.7in,4.0in]{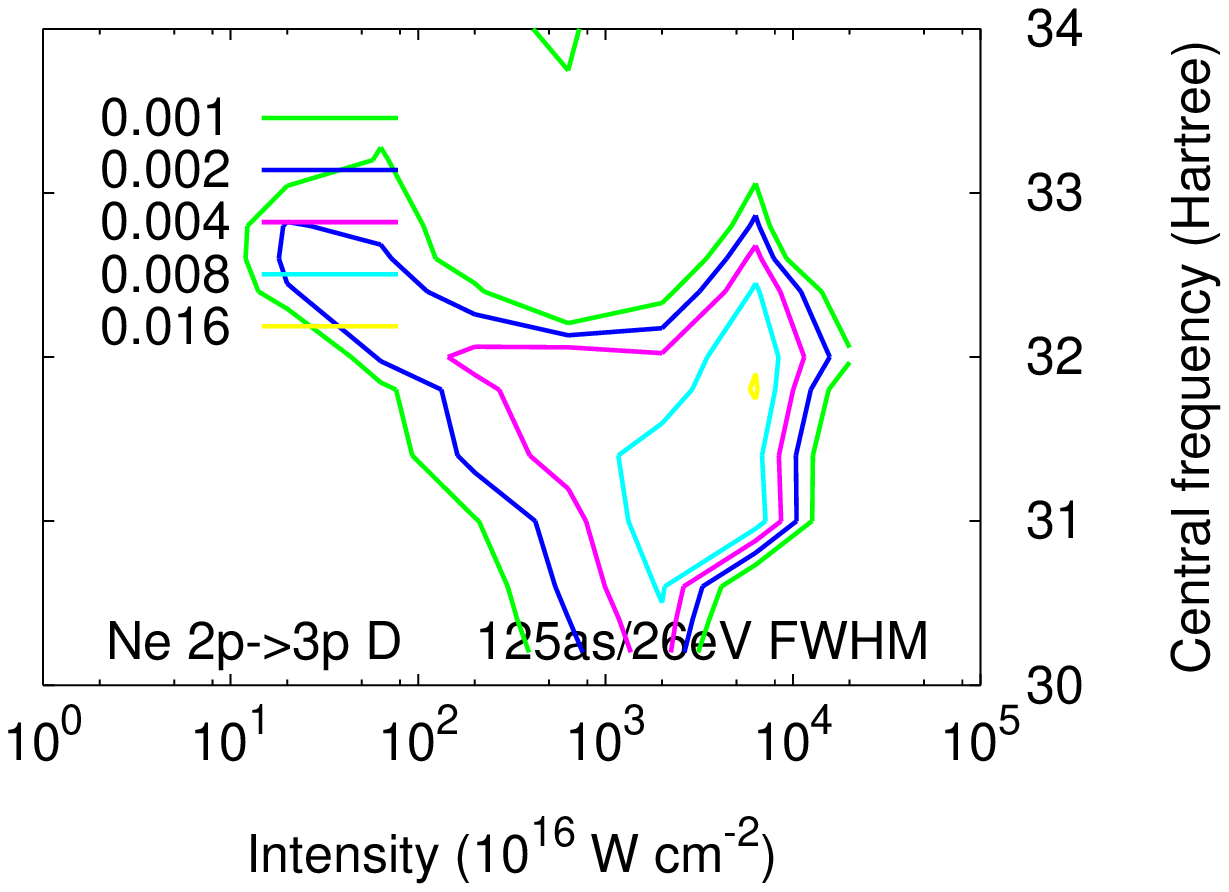}} \\
\resizebox{0.6\columnwidth}{!}{\includegraphics*[0.8in,0.4in][5.7in,4.0in]{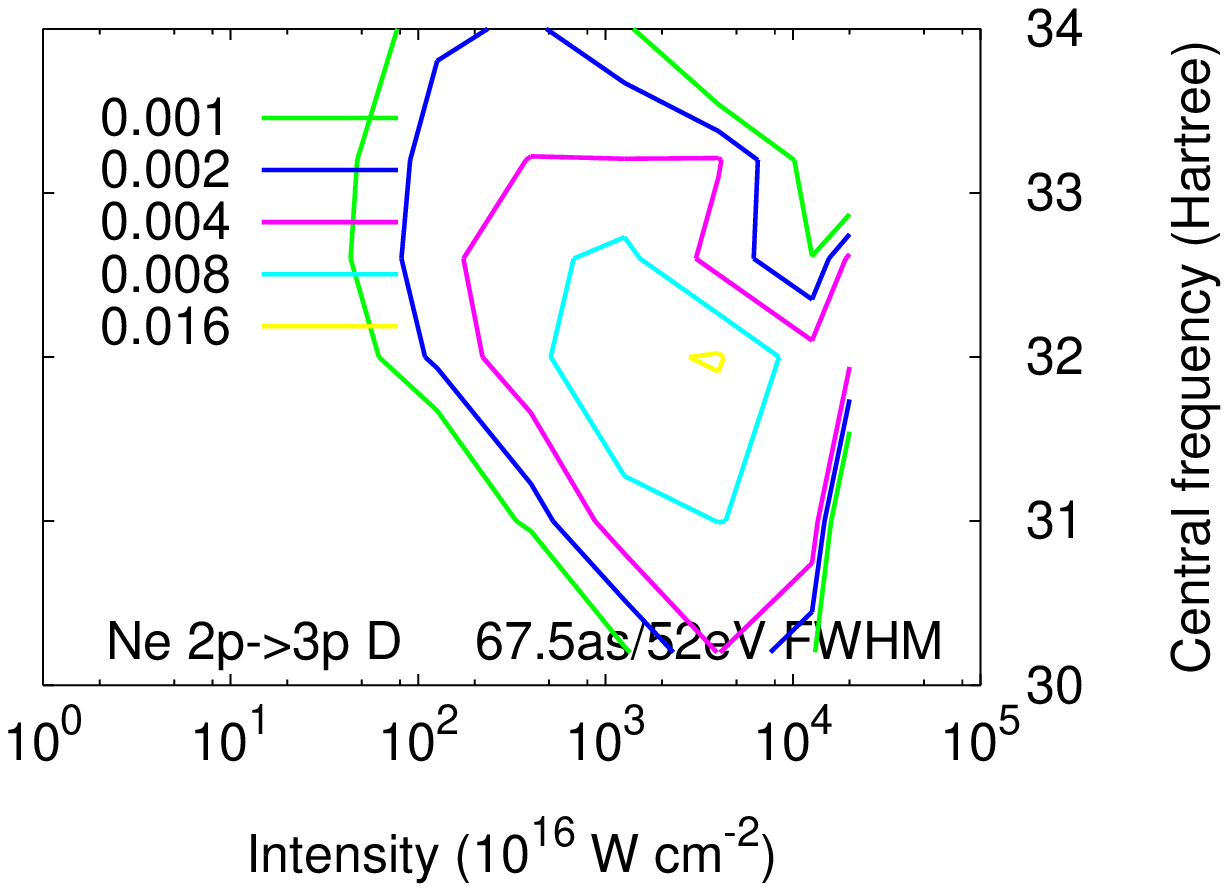}} \\
\end{tabular}
\caption{(Color online) Populations calculated for the 2p$^{-1}$3p D state in Neon.
\label{neon_popfig}}
\end{figure}

The excitation energies of Neon are large due to its closed-shell structure.  The first excitation energy of Neon is approximately
16.7eV~\cite{nist}, and so we would expect a pulse with FWHM 16.7 $\times$ 118\% = 19.6eV to be optimal.  We find that
the population transfer is dominated by transitions to the 2p$^{-1}$3p D state, with population transfer to other states 
about an order of magnitude lower in magnitude.

For much smaller bandwidths and longer durations, we calculate population transfer to this D state 
that clearly indicates the presence of
the two sweet spots.  In the top two panels of Fig.~\ref{neon_popfig}, at 500as/6.5eV FWHM and 250as/13eV FWHM, the
two sweet spots are clearly visible.  The conventional-wisdom
sweet spot occurs as expected around 32.4 Hartree in the top panel at 500as/6.5eV FWHM, but it 
shrinks in size and is overwhelmed by the
red-detuned sweet spots which becomes dominant in the lower figures at shorter duration.

By fitting
the results in Fig.~\ref{neon_popfig} to a quadratic function of bandwidth and fluence, we find that the global optimum
for population transfer to the 2p$^{-1}$3p D state in Neon via impulsive Raman near the 1s K-edge of occurs at approximately
82as FWHM (39.6eV FWHM), 31.9 Hartree (868eV) central frequency, and
8 $\times$ 10$^{19}$ W cm$^{-2}$ intensity.
Optimum population transfer is obtained with shorter pulses than would be expected.

In summary, for impulsive Raman transitions in Neon we find that the red-detuned sweet spot is strongly dominant and that the
optimum population transfer occurs for pulses even shorter (with even greater bandwidth) than would be expected.  Very little population
transfer is obtained via the near-edge fine structure below K-edge, in the conventional-wisdom sweet spot.


\section{Conclusion}

For impulsive X-ray Raman excitation of valence states of atoms and molecules, 
is commonly assumed~\cite{tiger, mukamel2013} that transitions via discrete states are most effective in driving the
process, and that that significant population transfer may be driven at second order in the field intensity.
However, the results presented here demonstrate that higher-order behavior must be accounted for to describe
population transfer at high intensity.  We find that higher-order behavior may lead to optimum population transfer either at 
much longer duration, as it does for Lithium via the conventional-wisdom, near-edge sweet spot, or at much shorter duration,
as it does for Neon via the red-detuned sweet spot, than expected from second-order perturbation theory.
In figure work we will consider the merits of using either the conventional-wisdom, near-edge sweet spot, or the red-detuned,
high-intensity sweet spot identified in this work, for applications in multidimensional X-ray spectroscopy~\cite{mukamel2013}.

\section{Acknowledgments}

Calculations have been performed on the Lawrencium supercluster at Lawrence Berkeley National
Laboratory (LBNL) under the support of the Laboratory Research Computing program,
http://scs.lbl.gov.
Matthew Ware is supported by the Stanford Graduate Fellowship.
Dan Haxton is supported by the Peder Sather Grant Program. 
This research is further supported 
by the U.S. Department of Energy, Office of Basic Energy Sciences, 
Chemical Sciences, Geosciences, and Biosciences Division:
through the Stanford PULSE Institute at the 
SLAC National Accelerator Laboratory, and through Lawrence Berkeley
National Laboratory, US DOE contract DE-AC02-05CH11231.

\bibliography{no2bib}

\end{document}